# Optical time domain reflectometry with low noise waveguide-coupled superconducting nanowire single-photon detectors

C. Schuck, W. H. P. Pernice, X. Ma, and H. X. Tang[*]

*Department of Electrical Engineering, Yale University, New Haven, CT 06511, USA*

We demonstrate optical time domain reflectometry over 200 km of optical fiber using low-noise NbTiN superconducting single-photon detectors integrated with $Si_3N_4$ waveguides. Our small detector footprint enables high timing resolution of 50ps and a dark count rate of 3 Hz with unshielded fibers, allowing for identification of defects along the fiber over a dynamic range of 37.4 dB. Photons scattered and reflected back from the fiber under test can be detected in free-running mode without showing dead zones or other impairments often encountered in semiconductor photon-counting optical time domain reflectometers.

* email: hong.tang@yale.edu

Optical time domain reflectometry (OTDR) is an efficient, nondestructive technique to diagnose the physical condition of an optical fiber *in situ* [1,2,3]. By launching laser pulses into the fiber and detecting the returning light from reflecting and scattering sites, it is possible to get information about attenuation properties, loss and refractive index changes in the fiber-under-test (FUT) [4]. Defects in the fiber-link can be localized with high spatial resolution over distances of more than a hundred kilometers by analyzing the returning optical signal in the time-domain. The achievable measurement range and two-point resolution of an OTDR system crucially depend on the detector used to monitor the weak backscattered signal. With increasing distance the light scattered or reflected along the FUT suffers from stronger attenuation and eventually reaches the detector noise level. Hence, the sensitivity of an OTDR system is determined by the noise equivalent power (NEP) of the detector which ultimately limits the measurement range.

Most commercial fiber-link characterization systems employ linear detectors, for example, p-i-n or avalanche photodiodes. This mature technology is well suited for in-field measurements. However, conventional OTDR systems are fundamentally limited by the bandwidth dependence of the NEP for linear photodetectors: high two-point resolution requires high receiver bandwidth which in turn reduces the OTDR sensitivity because the detector (amplifier) noise is proportional to the square-root of its bandwidth [5,6]. The competitive relation between resolution and measurement range therefore critically limits their suitability for monitoring an increased number of passive optical components and longer fiber distances with high resolution as, e.g., in next generation optical access networks.

Higher sensitivity can be achieved in photon-counting (ν) OTDR, employing single-photon detection techniques. The main advantage of ν-OTDR systems over their conventional counterparts stems from the lower NEP of single-photon detectors as compared to linear detectors, resulting in larger dynamic range and higher two-point resolution [3,6]. Such ν-OTDR schemes have been demonstrated with InGaAs/InP avalanche photo diodes (APD) operated in Geiger-mode [3,5], using silicon photon-counting modules in combination with telecom to visible frequency up-conversion [7,8], and with nanowire-meander superconducting single photon detectors (SSPD) [9,10]. APDs suffer significantly from various detection noise mechanisms, namely, afterpulsing [3], charge persistence [5] and memory effects [11], which degrade the signal and result in dead zones. Their use in ν-OTDR systems therefore relies on (rapid) gating and complex signal control systems. SSPDs on the other hand can be operated in free running mode and combine low NEP at telecom wavelengths with high timing resolution [12,13,14], which has previously been exploited for quantum key distribution [15] and time-of flight ranging [16,17].

Here, we demonstrate a ν–OTDR system using a low-noise waveguide-coupled superconducting nanowire single-photon detector in travelling wave geometry. We achieve low dark count rates by fabricating niobium titanium nitride (NbTiN) nanowires with a minimized footprint directly on top of a waveguide.

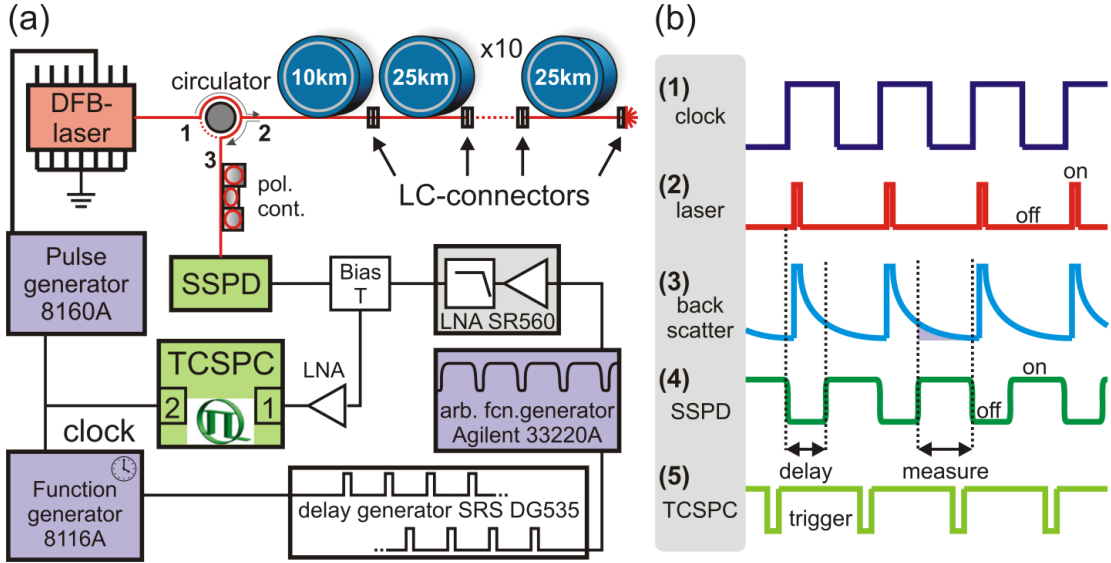

Fig. 1 (a) OTDR setup. Pulses from a DFB Laser are launched into the FUT consisting of 11 spools of SMF-28 fiber. The signal reflected back from the FUT is coupled out with a circulator and detected with a low noise NbTiN nanowire SSPD. For investigation of particular FUT-sections, the bias current through the nanowire SSPD can be switched on and off. The corresponding time trace is programmed with an arbitrary function generator and the relative gating window with respect to the laser pulses is adjusted with a delay generator. The bias current is supplied from a battery powered low noise amplifier (LNA). The output pulses from the SSPD are amplified and fed into a time-correlated single-photon counting unit (TCSPC). (b) A reference clock (1) synchronizes the laser pulses (2), the backscattered signal (3), the SSPD gating (4), and the TCSPC data acquisition system (5).

NbTiN is chosen for its low noise SSPD performance [18,19,20]. Travelling wave detectors achieve high detection efficiency for telecom wavelength photons [14,21] despite reduced nanowire dimensions as compared to traditional meander-type SSPDs [12,13]. Operation in free-running mode allows us to diagnose more than 200 km of optical fiber. The observed OTDR trace shows no artificial features or dead zones.

Our photon-counting OTDR setup is shown in figure 1 (a). Laser pulses are launched into the FUT via ports 1→2 of a circulator which couples the backscattered photons out via ports 2→3 and guides them towards the NbTiN nanowire SSPD housed in a liquid helium cryostat. The FUT consists of one 10.6 km spool of bare SMF-28 fiber with FC/APC connectors followed by ten 25.3 km SMF-28 fiber spools with LC/PC-connectors. All fibers are connected in series with standard mating sleeves. To identify the last spool after 263 km of fiber in an OTDR measurement the final LC-connector is left un-terminated to cause reflection at the glass-to-air interface (up to 4%).

For OTDR applications, it is desirable to have a laser source supplying short pulses of high power at a user-defined repetition rate. Given a detector of sufficient timing accuracy, the achievable two-point resolution is then determined by the laser pulse length. High laser power results in a larger number of photons scattered back towards the detector and thus reduces the data acquisition time needed to achieve a given OTDR measurement range. To unambiguously identify defects in the fiber, it is furthermore necessary to adjust the pulse repetition rate $R_{rep}$ to the total length of the FUT, $L_{tot}$. Overlapping echoes from two consecutive pulses are avoided when $R_{rep}^{-1} \leq 2L_{tot}c_f$, where $c_f = 4.9\,\text{ns}/\text{m}$ is the propagation delay for telecom wavelength photons in a fiber. Here, we generate customized pulses of width $\tau_p = 50$ ns and adjustable repetition rate from a DFB laser diode (SEI, SLT5413) supplying 10.5 mW of pulse peak power at 1550 nm wavelength. The repetition rate is set using a function generator (HP8116A) which acts as a clock to the whole OTDR system, i.e. laser, SSPD and data acquisition system (see Fig. 1 (a) and (b)). To maximize the OTDR measurement sensitivity, we adjust the clock rate to 300 Hz, exceeding the corresponding 263 km length of the fiber-under-test such that photons reflected from the open fiber end are able to travel back to the circulator before the next pulse is launched. The laser pulse width and height are set with a pulse generator (HP8160A) in burst mode, which is triggered by the clock frequency. The output of the pulse generator is applied directly to the laser diode cathode, with the anode set to ground. Importantly, this allows us to slightly reverse-bias the diode to switch it completely off in between pulses. Otherwise, any background light (even the diode's incoherent spontaneous emission below the lasing threshold) would severely compromise the measurement of the weak backscattered signal from the FUT.

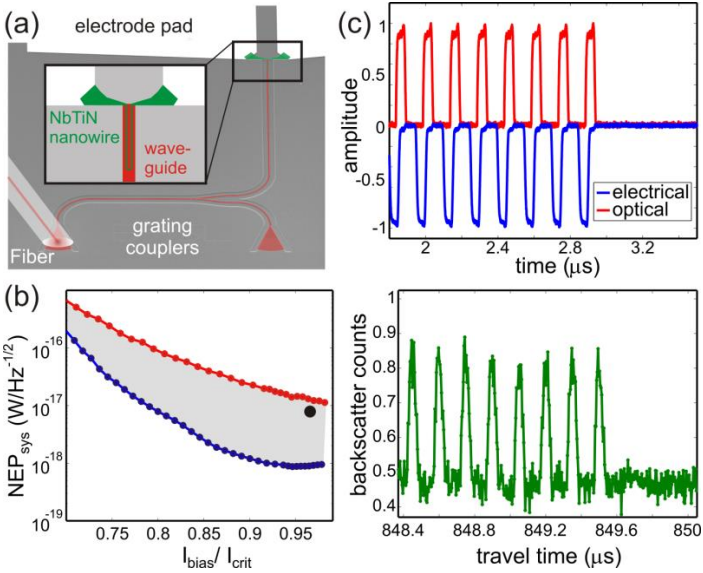

Fig. 2. (color online) (a) False-color SEM image of an integrated nanowire SSPD (green) on a SiN photonic waveguide (red). The 40 $\mu m$ long U-shaped NbTiN nanowire is patterned directly on top of a 330 $nm \times 1\ \mu m$ SiN waveguide (inset) and absorbs photons in the waveguide along their direction of propagation. (b) System noise equivalent power ($NEP_{sys}$) as a function of bias current ($I_{bias}$) in units of critical current ($I_{crit}$), taking into account the fiber-to-waveguide coupling efficiency for daylight conditions (red) and minimal ambient light conditions (blue). The black dot represents the conditions during the ν-OTDR measurements. (c) Output of the pulse generator in burst mode (blue) applied to the DFB laser diode (50 ns width, 150 ns period). Optical output pulses as detected with a fast photodiode (red). Backscattered signal from the FUT at around 111 km as detected with the SSPD after 10 min data acquisition (green, bottom) showing a spatial resolution of approximately 10 m.

The backscattered photons from the FUT are coupled out at circulator port 3 and guided to the travelling wave NbTiN superconducting nanowire single-photon detector inside the cryostat at 1.6 K. Coupling of light from the optical fiber into the on-chip photonic waveguide is achieved with an optical grating coupler. The SSPD is realized as an 8 nm thin, 75 nm wide U-shaped NbTiN-nanowire of 40 μm length on top of a 1 μm wide SiN photonic waveguide [22], see Fig. 2 (a). In this traveling wave geometry [21], approximately 95% of all photons in the waveguide are absorbed by the NbTiN-nanowire [22], resulting in a detection efficiency of 53% at telecom wavelengths when biased close to the critical current of the wire [20]. Accounting for the photon coupling loss from the fiber into the SiN waveguide, we obtain a maximal system detection efficiency of 4.3%. Importantly for OTDR measurements, SSPDs fabricated from NbTiN (rather than the more commonly used NbN) have been found to exhibit attractive low noise characteristics [18]. For the NbTiN nanowire SSPD used here, we find a dark count rate of less than 10 Hz over the entire bias current range, mainly limited by stray light [20]. The resulting system $NEP_{sys} \approx 10^{-17} - 10^{-18}\ W/\sqrt{Hz}$ close to the critical current is dependent on ambient light conditions, see Fig. 2 (b). For an OTDR measurement in daylight conditions, we hence expect to operate rather at the upper end of the NEP-range (i.e. $10^{-17} W/\sqrt{Hz}$) since ambient light leaks into the bare fiber and is efficiently guided to the on-chip detector.

The SSPD could, in principle, be operated continuously in free running mode. However, we find that the (residual) circulator transmission from port 1→3 is sufficient to cause a large number of detection events which exceed the weak backscattered signal by far. The light bypassing the FUT can cause a large amount of undesired detection events and even drive the SSPD into the normal state, where it is insensitive to photon absorption. We, therefore, reduce the bias current in synchronization with the clock frequency to switch the SSPD completely off for the short time when the laser pulses pass through the circulator. For the remaining time, we keep the bias current at approximately 95% of the critical current in free-running mode, thus detecting the OTDR signal trace of photons reflected back from anywhere within the FUT (traces (3) and (4) of Fig. 1 (b)). The clock signal is derived from the same reference used to trigger the laser. A delay generator (SRS DG535) allows us to precisely switch off the detector with respect to the launch time of the laser pulse (see traces (1), (2) and (4) in Fig. 1 (b)). The off period of the detector is set with an arbitrary function generator (Agilent 33220A) also triggered on the clock signal. The shape of the corresponding output pulses is designed to suppress transient oscillations of the bias current (which would exceed the nanowire critical current during the on-off switching). A battery powered low noise amplifier (SR560), which also acts as a low pass filter, supplies the bias current to the SSPD via a 100 kΩ resistor and a bias-T (ZFBT-4R2G+). We thus effectively reset the SSPD at the clock frequency and are able to operate it in free running mode for the rest of the clock cycle. Note that the main purpose of the gating is only to switch the detector off during the time when the laser pulse has not yet reached the FUT. In addition, this gating can also be used to investigate only particular sections of the FUT by adjusting the delay and on-time window.

The SSPD output is amplified with high-bandwidth low noise amplifiers and fed into a time correlated single-photon counting system (TCSPC, PicoHarp 300). This TCSPC unit is operated in time-tagged time-resolved (TTTR) mode recording all detection events with 4 ps resolution. While this suggests sub-centimeter precision, defect localization is ultimately limited by the timing accuracy of the counting system jitter of approximately 50 ps [20]. To extract the temporal delay of each detection event with respect to the launch time of the laser pulses, we use the second channel of the TCSPC unit to record the pulses derived from the clock frequency (see Fig.1 (b), traces (1) and (5)). We then create list-files of all arrival times for both channels from which we calculate the time delay $\Delta t$ between photon detection events in channel 1 with respect to the clock signal recorded in channel 2. This time delay translates to the distance $\Delta s = \Delta t/(2c_f)$ which the photon travels before being scattered or reflected back. Since the SSPD is operated in free running mode while a pulse is propagating in the FUT, the entire OTDR trace is reconstructed by calculating the waiting time distribution recorded with the TCSPC unit.

The resulting time trace of backscattered photons detected with our low-noise SSPD is shown in figure 3 as a function of fiber length. To reduce the total amount of generated data in the list-files, we acquired the trace in two steps. We first perform an OTDR measurement of the initial 0-120 km of fiber by adjusting the SSPD bias current and repetition rate accordingly. For the measurement of the remaining fiber stretch, we then set the repetition rate to cover all 263 km of fiber. In this case, we program the arbitrary function generator to only switch the SSPD to the high bias current regime after photons from the first approximately 80 km have already passed. This allows us to increase the measurement sensitivity by averaging over longer times, omitting the vast amount of detection events originating from reflections in the initial part of the fiber which are already accounted for by the first (shorter) measurement. Both traces are then matched in the region where they overlap (86-111 km). Since they were acquired with respect to the same clock signal, the backscatter features nicely overlap as shown in the inset of Fig. 3 for a section at 86 km.

The OTDR trace exhibits a number of peaks caused by Fresnel reflections from the refractive index change at the tiny air gap between two fibers at each fiber connector. The first 10.6 km fiber spool followed by at least eight 25.3 km spools can by clearly identified by means of the reflection peaks before the noise level is reached, as shown in Fig. 3. We attribute the different peak heights to the fact that some connectors are better mated than others. Despite the noise at the end of the measurement range, the strong reflection from the glass to air transition at the open fiber end after 263 km (eleven fiber spools) is still visible in our OTDR measurement. The OTDR data also allow us to extract the round-trip attenuation due to Rayleigh scattering in the fiber-under-test from the linear slope of the trace. The attenuation of each fiber spool can be obtained by fitting the slope of the corresponding part of the OTDR trace. As an example, in Fig. 3 we show the exponential fit to the data of the fifth spool (86-111 km) for both measurements (0-120 km and 80-270 km) yielding an attenuation of 0.196 dB/km in either case. To extract the rms noise level observed as the tail of the OTDR trace in Fig. 3, we use the data in the 233-263 km region, where the contribution from backscattered photons is negligible. The dark count rate during the OTDR measurement can be estimated by integrating the absolute number of (dark) counts per second in the last 30 km of fiber and extrapolation to a full laser pulse cycle. We thus obtain a dark count rate of 3 Hz corresponding to an NEP of approximately $8 \times 10^{-18} W/\sqrt{Hz}$ at a bias current of 95% of the critical current, consistent with the independently determined NEP shown in Fig. 2 (b). From the intersection of the slope with the rms noise level at 191 km of fiber, we then determine the measurement range of our ν–OTDR system as 37.4 dB.

The data in figure 3 was acquired with laser pulses of 50 ns width, which sets the limit of the achievable two-point resolution to about 10 m Since dispersion effects are negligible for such long pulses, we observe that the width (50 ns) of the reflection peaks (see Fig. 3 inset, for example) does not change over the entire measurement range. A train of such 50 ns pulses with a period of 150 ns is shown in figure 2 (c). The counts were accumulated in 10 minutes and all pulses in the OTDR trace remain clearly discernible as they are derived from the pulse generator signal applied to the laser diode. We find no evidence of dead zones or other measurement artifacts (e.g. from afterpulses) commonly encountered in time-gated applications using APDs. In order to achieve higher resolution, shorter pulses are desirable since the detector timing jitter (50 ps) essentially permits localization of defects with down to 1 cm accuracy [23]. The shortest optical pulses that could be generated from the DFB laser in our setup were 15.4 ns long, corresponding to a two-point resolution of about 3 m in optical fiber. However, due to a reduced pulse amplitude in this regime we increased the pulse duration to 50 ns, where the maximum pulse power of 10.5 mW was reached. For the SMF-28 fiber used here (zero-dispersion length $\lambda_0$=1313 nm, typical zero dispersion slope $S_0$=0.086 ps/(nm²·km)) a two-point resolution of ≤10 cm is achievable over 200 km length using 100 ps pulses with our low-jitter SSPD.

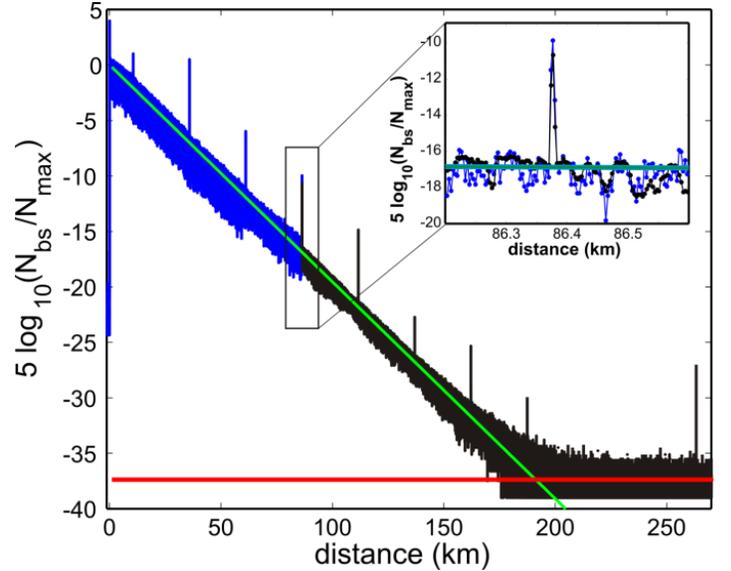

Fig. 3. (color online) OTDR traces of the backscattered photons ($N_{bs}$) measured using laser pulses of 50 ns width. Each peak is associated with the Fresnel reflection of the laser pulse from the connector pair of adjacent fiber spools. The reflection from the open fiber end after 263 km is strong enough to rise above the noise level. The inset shows the overlapping reflections from the connector at 86 km for the two measurements (0-120 km: blue, 80-270 km: black). The rms noise level (red) was calculated from the data at the tail of the OTDR trace (233-263 km). The measurement range extracted from the intersection of the slope with the rms noise level is 37.4 dB.

In conclusion, we have demonstrated photon-counting OTDR with superconducting nanowire single photons over a measurement range of more than 200km in telecom optical fiber. OTDR data are acquired with high resolution by operating the SSPD in free-running mode during pulse propagation in the fiber-under-test. The increased OTDR measurement range as compared to conventional [3] as well as many other photon-counting OTDR implementations [5,7-10] is a consequence of the low noise equivalent power of our NbTiN-nanowire SSPD. Note that the SSPD used here features high detection efficiency for photons traveling in on-chip photonic waveguides. However, the system detection efficiency is reduced by the fiber-to-waveguide coupling loss which could, for example, be improved with somewhat more involved grating coupler designs or inverted tapers [24]. Higher coupling efficiency then directly translates to a lower NEP and in turn increased dynamic range of the OTDR system. The spatial resolution of 10 m (3 m over shorter distance) for our OTDR system was only limited by the pulse width of 50 ns (15 ns) achievable for the laser used here. Due to the high timing accuracy (50 ps) of our SSPDs [20], we anticipate that defects in a fiber can be localized with ≤10 cm precision when using laser systems with sub-nanosecond pulse duration. The full potential of our low-noise nanowire SSPD for OTDR applications can also be assessed by combining detection efficiency η, dark count rate D, and timing jitter $\Delta t$, into one figure of merit $H = \frac{\eta}{D \cdot \Delta t} = 2.9 \times 10^8$, which compares favorably with other detector technologies [12]. While the our SSPD is operated in a liquid helium cryogenic systems, which is not always a viable choice in OTDR measurements, closed cycle refrigerators offer an attractive alternative for operating SSPDs in a more rugged environment [25]. The combination of low noise equivalent power and high time resolution thus makes SSPDs a promising choice for fulfilling the requirements of next generation OTDR applications.

H.X.T acknowledges support from a Packard Fellowship in Science and Engineering and a CAREER award from the National Science Foundation. W.H.P. Pernice acknowledges support by the DFG grant PE 1832/1-1. We want to thank Dr. Michael Rooks and Michael Power for their assistance in device fabrication and Dr. Robin Cantor for substrate preparation.

References

1. M. K. Barnoski, S. M. Jensen, Appl. Optics **15**, 2112 (1976).
2. M. K. Barnoski, M. D. Rourke, S. M. Jensen, R. T. Melville, Appl.Optics **16**, 2375 (1977).
3. P. Eraerds, M. Legré, J. Zhang, H. Zbinden, N. Gisin, J. Lightw. Technol. **28**, 952 (2010).
4. D. Derickson, Ed., Fiber Optic Test and Measurement, Prentice Hall (1998).
5. M. Wegmuller, F. Scholder, N. Gisin, J. Lightw. Technol. **22**, 390 (2004).
6. P. Healy, Opt. Quant. Electron. **16**, 267 (1984).
7. E. Diamanti, C. Langrock, M. M. Fejer, Y. Yamamoto, H. Takesue, Opt. Lett. **31**, 727 (2006).
8. M. Legré, R. Thew, H. Zbinden, N. Gisin, Opt. Express **15**, 8237 (2007).
9. J. Hu, Q. Zhao, X. Zhang, L. Zhang, X. Zhao, L. Kang, P. Wu, J. Lightw. Technol. **30**, 2583 (2012).
10. M. Fujiwara, S. Miki, T. Yamashita, Z. Wang, M. Sasaki, Opt. Express **18**, 22199 (2010).
11. A. Dalla Mora, D. Contini, A. Pifferi, R. Cubeddu, A. Tosi, F. Zappa, Appl. Phys. Lett. **100**, 241111 (2012).
12. R. H. Hadfield, Nat. Photon. **3**, 696 (2009).
13. G. N. Gol'tsman, O. Okunev, G. Chulkova, A. Lipatov, A. Semenov, K. Smirnov, B. Voronov, A. Dzardanov, C. Williams, R. Sobolewski, Appl. Phys. Lett. **79**, 705 (2001).
14. W. H. P. Pernice, C. Schuck, O. Minaeva, M. Li, G. N. Goltsman, A. V. Sergienko, H. X. Tang, Nat. Comm. **3**, 1325 (2012).
15. H. Takesue, S. W. Nam, Q. Zhang, R. H. Hadfield, T. Honjo, K. Tamaki, Y. Yamamoto, Nat. Photon. **1**, 343 (2007).
16. R. E. Warburton, A. McCarthy, A. M. Wallace, S. Hernandez-Marin, R. H. Hadfield, S. W. Nam, G. S. Buller, Opt. Lett. **32**, 2266 (2007).
17. M. G. Tanner, S. D. Dyer, B. Baek, R. H. Hadfield, S. W. Nam, Appl. Phys. Lett. **99**, 201110 (2011).
18. S. N. Dorenbos, E. M. Reiger, U. Perinetti, V. Zwiller, T. Zijlstra, T. M. Klapwijk, Appl. Phys. Lett. **93**, 131101 (2008).
19. S. Miki, M. Takeda, M. Fujiwara, M. Sasaki, A. Otomo, Z. Wang, Appl. Phys. Express **20**, 075002 (2009).
20. C. Schuck, W. H. P. Pernice, H. X. Tang, Sci. Rep. **3**, 1893 (2013).
21. X. Hu, C. W. Holzwarth, D. Masciarelli, E. A. Dauler, K. K. Berggren, IEEE Trans. Appl. Supercond. **19**, 336 (2009).
22. C. Schuck, W. H. P. Pernice, H. X. Tang, Appl. Phys. Lett. **102**, 051101 (2013).
23. O. V. Minaeva, A. Fraine, A. Sergienko, A. Korneev, A. Divochiy, G. Goltsman, Frontiers in Optics, Optical Society of America (2012).
24. D. Taillaert, P. Bienstman, R. Baets, Opt. Lett. **29**, 2749 (2004).
25. C. M. Natarajan, M. G. Tanner, R. H. Hadfield, Supercond. Sci. Technol. **25**, 063001 (2012).